\documentclass[twocolumn,prb,showpacs,superscriptaddress]{revtex4}
\usepackage[dvips]{graphicx}
\usepackage{amsmath,bm}
\usepackage[dvips]{color}
\usepackage{graphicx}
\usepackage{epsfig}

\begin{document}
\title{Hyperfine interactions in silicon quantum dots}
\author{Lucy V. C. Assali}
\affiliation{Instituto de F\'{\i}sica, Universidade de S\~ao Paulo,
CP 66318, 05315-970, S\~ao Paulo, SP, Brazil}
\author{Helena M. Petrilli}
\affiliation{Instituto de F\'{\i}sica, Universidade de S\~ao Paulo,
CP 66318, 05315-970, S\~ao Paulo, SP, Brazil}
\author{Rodrigo B. Capaz}
\affiliation{Instituto de F\'{\i}sica, Universidade Federal do
Rio de Janeiro, Caixa Postal 68528, 21941-972 Rio de Janeiro, Brazil}
\author{Belita Koiller}
\affiliation{Instituto de F\'{\i}sica, Universidade Federal do
Rio de Janeiro, Caixa Postal 68528, 21941-972 Rio de Janeiro, Brazil}
\author{Xuedong Hu}
\affiliation{Department of Physics, University at Buffalo, SUNY,
Buffalo, New York 14260-1500, USA}
\author{S. Das Sarma}
\affiliation{Condensed Matter Theory Center, Department of Physics,
University of Maryland, College Park, Maryland 20742-4111, USA}
\begin{abstract}
A fundamental interaction for electrons is their hyperfine interaction (HFI) with nuclear spins.  HFI is well characterized in free atoms and
molecules, and is crucial for purposes from chemical identification of atoms to trapped ion quantum computing. However, electron wave functions
near atomic sites, therefore HFI, are often not accurately known in solids. Here we introduce an all-electron calculation for conduction
electrons in silicon and obtain reliable information on HFI. We verify the outstanding quantum spin coherence in Si, which is critical for
fault-tolerant solid state quantum computing.
\end{abstract}
\date{\today}
\maketitle


\section{Introduction}

The current interest in silicon in the context of spintronics and spin quantum computation \cite{Zutic_RMP04,DasSarma_SSC04,Kane_Nature98} arise
naturally from its perceived technological and fundamental advantages.  Technological advantages included scalability and unique material
processing capabilities, developed primarily by the well established and, already for half-a-century dominant Si microelectronics industry. The
fundamental advantage, which is the key motivation for our present work, is the very low abundance of finite-spin isotope ($^{29}$Si) in natural
Si, leading to very long electronic spin coherence times and thus enabling fault-tolerant spin qubit operations. Recent ESR experiments on
phosphorus donor electrons in isotopically purified Si show that electron spin coherence times are longer than 60 ms \cite{Lyon_JPCM06}, and
there have been tremendous recent experimental progress in the study of the Si:P system.\cite{Abe_PRB10, Fuechsle_NN10, Morello_Nature10}
However, fabricating P donor arrays with the necessary atomic-scale precision \cite{KHD_PRL02} is a difficult challenge still to be overcome
\cite{Australia_Nano, Review2, Simmons_NanoL}. An alternative to the Si:P system uses electron spins in gate-confined Si quantum dots (QD) as
qubits. There have been extensive experimental studies of few-electron spin dynamics in gated semiconductor (in particular GaAs) QDs for the
past decade \cite{Hanson_RMP}, while recent experiments \cite{Wisconsin_PRB03, Liu_PRB08, Sandia_Preprint} have shown that single-electron QDs
are well within reach in Si as well \cite{Wisconsin_APL08}.

In Si:P, quantum coherence of {\it donor confined} electron spins is limited by their well-known hyperfine interactions (HFI) with the random
$^{29}$Si nuclei in the environment, whose dipolar interaction induced nuclear spin dynamics and the subsequent electron spin spectral diffusion
is the main source of donor electron spin decoherence.\cite{Witzel_SD,WHD_PRB07,Ivey_PRB75,Park_PRL09,Abe_PRB10,Witzel_PRL10,note_SiP} Clearly,
accurate knowledge of HFI would also be required to evaluate spin decoherence for an electron {\it confined in a QD} and to assess the magnitude
of the Overhauser field, i.e. the net field felt by the electron spin due to all the $^{29}$Si nuclear spins in a Si QD.  The strength of HFI in
a QD with smooth confinement could be directly derived from the HFI of an extra conduction electron in bulk Si.  In spite of its considerable
importance, however, these HFI parameters have never before been evaluated for Si {\it conduction electrons}, and our work fills this
information gap.

Motivated by spintronic and quantum computing considerations, we develop an ab initio study of the HFI parameters between Si conduction
electrons and $^{29}$Si nuclei. Our results are directly relevant to the prospective application of Si as the material of choice in these
applications. In addition, they are in good agreement with the existing experimental data related to
HFI.\cite{Shulman_PR56,Wilson_PR64,Dyakonov_PRB92} Our results provide the basis for a local spin density approximation (LSDA) for the HFI,
similar to LDA in electronic calculations. The LSDA is ideal for treating electrons in relatively extended QD states in Si, as illustrated
below.

We emphasize that the calculation of the HFI for a single extra electron in an unperturbed system, like bulk Si, introduces fundamental
methodological difficulties as compared to the spatially localized electron bound to shallow impurities in
semiconductors\cite{Ivey_PRB75,Park_PRL09,assali,larico,justo-fred} due to the extended nature of the bulk electron.  In general donor electron
wave functions cannot be expressed simply as an envelope function multiplied by the Bloch states of the conduction electrons.\cite{Kohn_review}
Therefore, even though calculation of HFI at an impurity site or at neighbouring sites for shallow impurities in semiconductors is
straightforward nowadays,\cite{Park_PRL09,assali,larico,justo-fred} we cannot easily extract information about the underlying Bloch functions,
which determine the conduction electron HFI.

The paper is organized as follows.  In Section II we introduce the theoretical method with which we perform the hyperfine calculation for bulk
Si.  In Section III we show the results on conduction electron hyperfine interaction in Si and compare with existing experimental measurements.
In Section IV we calculate the hyperfine interaction for confined electron in a Si quantum dot, and compare our results with that in GaAs
quantum dots.  Lastly, in Section V we draw our conclusions.

\section{Theory and Formalism}

The HFI hamiltonian has the general form
\begin{equation}
H_{\rm HF} = {\bm I} \cdot {\bm A} \cdot {\bm S} \,,
\end{equation}
where ${\bm S}$ and ${\bm I}$ are the electron and nuclear spin operators, and ${\bm A}$ is the HFI tensor. The tensor components for a nucleus
at ${\bm R}_{I}$ can be written in terms of an isotropic term $a$, known as Fermi contact term, and of an anisotropic dipolar traceless tensor
$b_{ij}$:~
$ A_{ij} = a \delta_{ij} + b_{ij}$.
The labels $i$ and $j$ refer to coordinates $x$, $y$, and $z$ taken here along the conventional fcc cubic cell edges. We consider Si with one
extra electron at a fixed conduction band minimum ${\bm k}_i$. The system has axial symmetry with respect to the $i$-axis, so that the traceless
anisotropic tensor is diagonal and expressed as $b_{ij} = diag\{2b, -b, -b\}$.  The hyperfine tensor ${\bm A}$ is thus completely defined by the
scalar hyperfine parameters $a$ and $b$, as described in Ref.~\onlinecite{vandewallle}. Explicit expressions for these parameters
\cite{vandewallle} are determined by the electron spin density,
\begin{equation}
\rho_{S}({\bm r}) = \rho_{\uparrow} ({\bm r}) - \rho_{\downarrow} ({\bm r}) \,.
\label{eq:spin-dens}
\end {equation}
In particular the Fermi contact interaction is proportional to the spin density at the nuclear site, $\rho_{S}({\bm R}_{I})$.  From $a$ and $b$,
given in energy units, one can obtain the equivalent magnetic field created by the nuclear spin acting on an electron spin.  Reliable
calculation of the HFI parameters require precise values of $\rho_{S}({\bm r})$ for all ${\bm r}$, in particular in the vicinity ${\bm r}
\approx {\bm R}_I$.

Our calculations, performed within the Density Functional Theory (DFT) framework,\cite{ks} involve the full-potential linearized augmented plane
wave method (FP-LAPW),\cite{singh} as embodied in the WIEN2k package.\cite{blaha}  This state-of-the-art all-electron methodology includes spin
polarization of the core and valence states, spin-orbit coupling, and relativistic effects.  No shape assumption (e.g. the usual spherical
constrain) to the potential is involved.  The generalized gradient approximation of Perdew, Burke and Ernzerhof (GGA-PBE) is used for the
calculation of the exchange-correlation potential.\cite{pbe} Calculations of HFI of negatively charged bulk Si are performed for several Si
supercell sizes, ranging from $N$ = 8 to 64 atoms, each with a single extra electron. In practice this corresponds to different average
electronic densities $\rho_N = 2e/(N \Omega_{PC})$, where $\Omega_{PC}$ is the volume of the Si primitive cell (PC), which contains two atoms.
A uniform positive jellium background with density $|\rho_N|$ is included to cancel the long range multipole interactions of charged
supercells.\cite{assali}  Grids from 8 to 125 k-points are used to sample the irreducible wedge of the Brillouin zone,\cite{mp} according to the
supercell size.  The calculation is performed self-consistently until convergence on the total energy and total charge is achieved.  This
approach realistically describes the electronic properties, and has previously been used to calculate HFI of impurities in
semiconductors.\cite{larico,justo-fred}  Our calculations give a Si lattice parameter of 5.47 \AA [experimentally measured value (exp.) is 5.43
\AA], bulk modulus $B = 90$ GPa (exp. 98.8 GPa), and cohesion energy $E = 4.55$ eV (exp. 4.63 eV).

We follow a two-step procedure to compute the hyperfine parameters at a given Si site for a supercell with an extra electron: 1) The converged
charge density is obtained from a standard spin-polarized self-consistent calculation. 2) This charge density is used to get the potential
energy of the Kohn-Sham equation, and the Hamiltonian is written assuming a fixed ${\bm k}_i$ corresponding to one of the six conduction band
minima.  In this way we constrain the extra electron to a fixed k-point, in analogy with the case of impurities in
semiconductors.\cite{assali,larico,justo-fred}  From the sum over all occupied bands $n$ of $|\phi_n^\sigma (\bm  r)|^2$ and of
$|\phi_n^{\bar\sigma} (\bm r)|^2$ for each spin orientation $\sigma$ and $\bar\sigma$ (where $\phi_n^{\sigma,\bar\sigma}$ are the Kohn-Sham
eigenstates), we obtain the net spin density [Eq.(\ref{eq:spin-dens})], as described in the appendix of Ref.~\onlinecite{larico}.   Note that
for neutral bulk Si $\rho_S = 0$, so the calculated $\rho_S (\bm r)$ is entirely due to the added electron. This procedure provides as good
quantitative estimates of spin density as possible within available theoretical treatments.

\section{Hyperfine Interactions for a Conduction Electron in Bulk Silicon}

\begin{figure}
{\centering\resizebox*{7.0cm}{!} {\includegraphics*{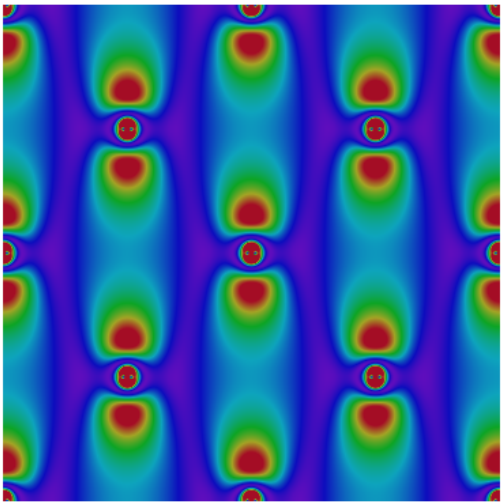}}} \caption{(Color online) Electron spin density in a (100) plane for
bulk Si with an extra electron in the conduction band minimum at $k_z$. The vertical axis is $z$ and the color scheme runs from red (high
density) to purple (low density), following the rainbow sequence. The circular high density spots are the Si atomic sites.} \label{fig1}
\end{figure}

Figure \ref{fig1} shows the spin density in the ($100$) crystal plane calculated with an 8-atom supercell for negatively charged Si, where the
extra electron is kept at ${\bm k} = {\bm k}_z$ at the conduction band edge.  Note the anisotropy of the distribution, elongated in the axial
$z$ direction. Comparison of this spin density with the charge density in the same plane for the conduction band ${\bm k}_z$
pseudo-wavefunction, given in Fig.~1(b) of Ref.~\onlinecite{Koiller}, illustrates the similarity of the two distributions, except for regions
just around the atomic sites.  This is because results in Ref.~\onlinecite{Koiller} were obtained within the pseudopotential approach, where the
core region potentials are very different from the actual ones. The same similarity is obtained in any other plane. These considerations, and
the fact that $\rho_S (\bm r)$ is nonzero with the added electron, allows us to identify  $\rho_S (\bm r)$ in our all-electron calculation with
the charge distribution of the extra electron, which in a one-electron scheme corresponds to $|\Psi_{k_i}{(\bm r)}|^2$, the absolute value of
the Bloch function squared.

\begin{figure}
{\centering\resizebox*{7.0cm}{!}{\includegraphics*{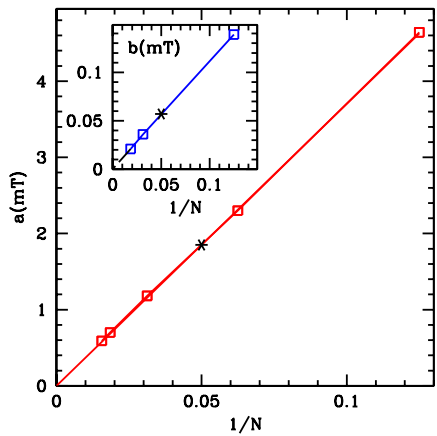}}} \caption{(Color online) Squares give the calculated hyperfine
parameters versus $1/N$, where $N$ is the number of Si atoms in the supercell. The asterisks give the interpolated values expected for natural
Si ($\sim 5\%$ of $^{29}$Si), with $a \approx 1.9$ mT and $b \approx 0.06$ mT. Solid straight lines are $a_N = (37/N)$ mT and $b_N = (1.1/N)$
mT, and give good fits to the data points.  These results are the basis of a local spin density approximation used in our quantum dot
calculations, similar to LDA in electronic calculations.} \label{fig2}
\end{figure}

Figure \ref{fig2} summarizes the calculated $^{29}$Si HFI as a function of $1/N$ in the negatively charged supercells.   Clearly, the HFI
parameters $a$ and $b$ scale linearly with the electronic density.  For natural Si, the isotropic parameter $a$ is of the order of a couple of
mT, while the anisotropic parameter satisfies $b/a \sim 3\%$ for all $N$.  We have also calculated the HFI parameters $a$ and $b$ for an $N=8$
supercell using the more recently developed GGA-WC exchange-correlation functional.\cite{Wu_PRB06}  The resulting HFI parameters show only minor
changes: $a$ ($b$) increases (decreases) by less than 5\% (8\%) as compared to the values shown in Fig.~\ref{fig2}.  Indeed, previous studies
comparing different exchange-correlation functionals showed that DFT methods are generally quite consistent with each other, and have at most
about 20\% systematic error compared to experimental observations.\cite{Adamo_JCP99}  We expect that for bulk Si, which contains no unsaturated
bonds, the systematic error should be smaller than 20\%.

\begin{figure}
{\centering\resizebox*{7.0cm}{!}{\includegraphics*{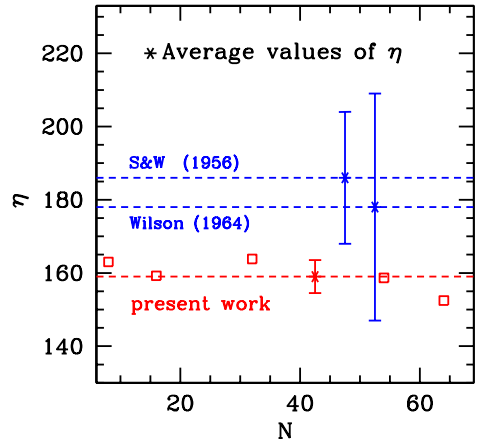}}} \caption{(Color online) Calculated $\eta$ for different supercell sizes (red
squares). The average value and standard deviation are given by the lowest (red) asterisk and respective error bar. Experimental results are
given by the upper (blue) horizontal dashed lines, following the average values of  Refs.~\onlinecite{Shulman_PR56} and
\onlinecite{Wilson_PR64}, labeled S\&W (1956) and Wilson (1964), respectively. The error bar for Wilson was obtained as an upper limit of the
error estimated from Ref.~\onlinecite{Wilson_PR64}, where $\eta_{\,\rm Si}$ is given in an expression related to $\eta_{\,\rm Ge}$. The
horizontal position of the asterisks is arbitrary, as they represent average values. The estimated value of $\eta \gtrsim 300$ in
Ref.~\onlinecite{Dyakonov_PRB92} is off the scale here.} \label{fig3}
\end{figure}

Within a mono-electronic framework, the relative weight of the conduction band wavefunction at a Si atomic site ${\bm R}_I=0$ may be quantified
by a parameter ($\eta$) defined below.\cite{Shulman_PR56}  It is widely adopted in experiments and in effective mass calculations.  The original
definition in a one-electron context and the analogous expression for the all electron framework here are, respectively
\begin{equation}
\eta_{\rm 1E} = \frac{|\Psi(0)|^2}{(|\Psi|^2)_{Av}} ~~~ \Leftrightarrow ~~~ \eta_{\rm AE}=\frac{\rho_S (0)}{(\rho_S)_{Av}}~,
\end{equation}
where in the denominators $(Q)_{Av}$ is the average of $Q$ over the normalization volume.  We omit the subscripts ``1E'' and ``AE'' below.
Ref.~\onlinecite{Shulman_PR56} inferred $\eta = 186 \pm 18$ for $^{29}$Si from NMR experiments. Later work \cite{Wilson_PR64} identified an
error in Ref.~\onlinecite{Shulman_PR56} and gave a corrected value of $\eta = 178 \pm 31$.  Ref.~\onlinecite{Dyakonov_PRB92} deduced $\eta$ by
extrapolating Overhauser shift data and obtained roughly twice the value reported in Ref.~\onlinecite{Shulman_PR56}, with $\eta \gtrsim 300$.
Our calculation constitutes the first theoretical estimate for $\eta$ and give $\eta = 159.4 \pm 4.5$.  The results are presented and compared
to the experimental estimates \cite{Shulman_PR56,Wilson_PR64} in Fig.~\ref{fig3}.

\section{Hyperfine Interactions for a Conduction Electron Confined in a Quantum Dot}

We now consider an electron confined in a QD near a [001] interface with a barrier material.  In this case the $\pm k_z$ valleys are lower in
energy and make up the electronic ground state.  By symmetry, $\rho_S$ is the same for $k_z$ and $-k_z$, so the results here apply to general
superposition states involving $\pm k_z$.  In the envelope function approach, the electron wave function in the QD is given by a bulk state
modulated by a slowly varying envelope function $F(\bm r)$, with $\int_{\rm all~space} |F|^2 d{\bf r} = 1$.  In the all-electron scheme, the
envelope modulation is $|F|^2$, since $|\Psi|^2$ is associated with $\rho_S$, while the bulk contribution to the spin density is normalized
within a PC, so that the overall spin density in a QD (normalized in all space) is
\begin{equation}
\rho_S^{QD} ({\bm r}) = |F({\bm r})|^2 {\Omega_{PC} }\rho_S^{PC}(\bm r) .
\end{equation}
In essence, our {\it ab initio} calculation for a supercell of volume $V$ is equivalent to having a uniform envelope function $F = 1/\sqrt{V}$
and a bulk spin density normalized in a PC.  In a QD, $F$ is non-uniform and extends over tens of nanometers.  However, this change does not
affect the calculation for the contact part of the interaction, because both the envelope function and the total spin density are normalized
over the QD.

The contact HFI field due to a nuclear spin at ${\bf R_I}$ in a QD is
\begin{equation}
a({\bf R}_I) = a_2 |F({\bf R}_I)|^2 \Omega_{PC} \,,
\end{equation}
where $a_2$ is extrapolated from our supercell calculations (see Fig.~\ref{fig2}) to a PC (a supercell with $N=2$).  Since we fit our
calculations as $a_N \approx (37/N)$ mT, $a_2 = 18.5$ mT. The $a({\bf R}_I)$ thus calculated is the interaction strength of the electron with a
particular nucleus at ${\bf R}_I$.  If we have a QD where every nucleus has a spin, with the nuclear spins arranged in a fully polarized state,
the total HFI $\cal A$ is a sum over the whole QD:
\begin{equation}
{\cal A} = \sum_{{\bm R}_I \in QD} a({\bm R}_I) = 2 a_2 \, ,
\end{equation}
where the factor $2$ accounts for contributions from the two nuclear sites per PC, and, to an excellent approximation, the envelope is taken to
be constant within each elementary PC.  The total HFI $\cal A$ corresponds to the magnetic field acting on the electron if all nuclear spins in
the QD are polarized.

In Table~\ref{table:OverhauserFields} we show estimates of the relevant Overhauser fields \cite{foot} and/or energy scales in Si and GaAs QDs.
The random Overhauser field $\delta \cal A$ for a high temperature nuclear reservoir is $\delta \cal A = \cal A/\sqrt{\cal N_S}$, with $\cal
N_S$ giving the total number of finite spin nuclei. The $T_2^*$ time is the electron spin dephasing time, given by $T_2^* = \hbar/\delta \cal
A$.  The first row of the table shows typical data for a GaAs \cite{Paget} QD that contains about $10^6$ nuclei, and is given as a benchmark for
comparison.  The 2nd row gives the data for a typical natural Si QD with $10^5$ nuclei (generally Si QDs are more strongly confined than GaAs
QDs).  With the three orders of magnitude difference in $\cal A$ in GaAs and natural Si, the random fields and $T_2^*$ are different by about
two orders of magnitude.  For an isotopically enriched $^{29}$Si sample, the difference goes down to one order of magnitude (3rd row); while for
an isotopically enriched $^{28}$Si sample (4th row), the difference goes up to three orders of magnitude.  The much smaller possible random
Overhauser field in Si QDs leads to much longer inhomogeneous broadening time $T_2^*$, ranging from 110 ns for a pure $^{29}$Si dot to 11 $\mu$s
for a 99.99\% purified $^{28}$Si dot. The long $T_2^*$ time indicates that in Si coherent electron spin manipulation is possible without spin
echoes. Conversely, if an inhomogeneous field is required to manipulate the electron spin states,\cite{Hanson_RMP, Petta_Science05, Zutic_RMP04}
in Si such a field probably needs to be applied.

\begin{table}
\begin{tabular}{|c|c|c|c|c|c|}
\hline host & $\cal N_T $ & $\cal N_S$ & $\cal A$  & $\delta {\cal A}$  &
$T_2^*$ \\
\hline
GaAs & $10^6$ & $10^6$ & 92 $\mu$eV (3.6 T) & 92 neV & 7.2 ns \\
Natural Si & $10^5$ & $5000$ & 210 neV (1.85 mT) & 3.0 neV & 0.22 $\mu$s \\
100\% $^{29}$Si & $10^5$ & $10^5$ & 4.3 $\mu$eV (37 mT) & 13.6 neV & 49 ns \\
0.01\% $^{29}$Si & $10^5$ & $10$ & 0.43 neV (3.7 $\mu$T) & 0.136 neV & 4.9 $\mu$s \\ \hline
\end{tabular}
\caption{Order of magnitude estimates of HFI that an electron can experience in a Si QD. The columns give the total number $\cal N_T$ of nuclei
in the dot, the number $\cal N_S$ of non-zero spin, the maximum $\cal A$ and random $\delta \cal A  = {\cal A}/\sqrt{\cal N_S}$ Overhauser
fields, when the nuclear spins are all aligned or in a random configuration, and the corresponding $T^*_2$ dephasing time.  The estimates for
GaAs are given for comparison.}
\label{table:OverhauserFields}
\end{table}

We also estimate the anisotropic HFI of a nuclear spin at ${\bm R}_I$ in the QD by including two contributions.  One is the near-field
contribution, dominated by the anisotropy of the spin density for ${\bm r} \approx {\bm R}_I$. Taking a small volume $\Omega \sim \Omega_{PC}$
around ${\bm R}_I$, containing $N_\Omega \sim 2$ Si atoms, the near-field contribution is $\Omega\, |F({\bm R}_I)|^2\, b_ {N_\Omega}$, and the
scaling $b_N=(1.1/N)$ mT can be used. The second is the far-field contribution due to nuclear spins randomly located inside the QD. Our
numerical estimates indicate that, for a QD with a radius of 20 nm, the near-field contribution is in the order of $10^{-4}$ Gauss, about 3\% of
the contact HFI, as expected. The far-field contribution is even smaller, at only $\sim 10^{-6}$ Gauss per nuclear spin. In short, in a Si QD
where a single electron has a smooth probability distribution, the anisotropic HFI is negligibly small.

\section{Conclusions}

In summary, we have performed a state-of-the-art all-electron calculation of hyperfine interaction in the Si conduction band.  Our study
introduces and validates an ab-initio approach to an open theoretical question of key relevance in spin behavior in a variety of materials, in
particular semiconductors, paving the way for such calculations in other crystalline systems.  Our calculated HFI strengths are consistent with
existing experimental observations.  The theoretical estimate for electron spin $T_2^*$ dephasing time in a natural Si QD is one to two orders
of magnitude longer than that in a GaAs QD.  The corresponding electron spin decoherence time should be at least two orders of magnitude longer,
formally and quantitatively verifying the advantages of Si as a host material for spin qubits.

\begin{acknowledgments}
This work was partially supported in Brazil by CNPq, CAPES, FAPESP, and FAPERJ.  BK, RBC, and HMP performed this work as part of the INCT
Program.  XH acknowledges support by NSA/LPS through ARO with grant number W911NF-09-1-0393.  SDS acknowledges support by NSA/LPS.
\end{acknowledgments}


\begin{thebibliography}{99}
\bibitem{note_SiP} Ultimately, donor electron spin coherence is limited by electron spin dipolar interaction between donors.\cite{Witzel_PRL10}
In other words, if P donor electron spins are used as qubits in a Si quantum computer, it is the qubit-qubit magnetic dipolar coupling, or qubit
cross talk, that sets the ultimate limit to the overall quantum coherence of the system.
\end{thebibliography}
\end{document}